\documentstyle[prd,12pt,aps,preprint,epsf]{revtex}

\begin{document}
\def\sqr#1#2{{\vcenter{\vbox{\hrule height.#2pt
        \hbox{\vrule width.#2pt height#1pt \kern#1pt
           \vrule width.#2pt}
        \hrule height.#2pt}}}}
\def\square{\mathchoice\sqr54\sqr54\sqr{6.1}3\sqr{1.5}6}
\tightenlines
\draft
\widetext
\title{Supersymmetric Quantum Mechanics for Bianchi Class A models }
\author{  J. Socorro{\footnote{E-mail: socorro@ifug2.ugto.mx}} and 
E. R. Medina{\footnote{E-mail: emedina@ifug3.ugto.mx}} \\
Instituto de F\'{\i}sica de la Universidad de Guanajuato,\\
Apartado Postal E-143, C.P. 37150, Le\'on, Guanajuato, Mexico.
}
\date{\today}
\maketitle

\begin{abstract}
In this work we present  cosmological quantum solutions for all Bianchi 
Class A cosmological models obtained by means of supersymmetric quantum 
mechanics . We are able to write one general expression for all bosonic 
components occuring in the Grassmann expansion of the wave function of 
the Universe for this class of models. These solutions are obtained by 
means of a more general ansatz for the so-called master equations.
\end{abstract}

\section{Introduction}
In recent years, many particular  solutions of the Wheeler-DeWitt (WDW) 
equation in standard quantum cosmology have been found  for Bianchi
Class A models \cite{OS}. In general, the wave functions  of these models 
have the form 
$\rm \Psi(q^\mu)= W(q^\mu) \, e^{\pm \Phi(q^\mu)}$, where $\rm \Phi$ is a 
solution to the corresponding classical Hamilton-Jacobi equation. One can
find similar solutions for a WDW equation that has been derived in the 
bosonic sector of the heterotic string \cite{Lidsey}.
Some time ago, Graham \cite{Graham1,Graham2} considered the WDW equation
in the ADM formulation of the Bianchi type IX model and found the
appropriate factorization in terms of the operators which may be considered
the {\it square roots} of this equation.  Graham was the first to provide a 
master equation for the Bianchi type IX models in Witten's supersymmetric 
quantum mechanics for the so-called auxiliary function 
$\rm f_\pm(q^\mu)$. However, he studied only the trivial case $\rm f=const$. 
Next, Obreg\'on {\it et al}, \cite{OSB} obtained ans solved the master
equation for the Bianchi type II models. For an application of the technique 
to the 2D Fokker-Planck equation see \cite{RSO}.  
 
Our main goal in this work, is to find the general form for all bosonic
functions that appear in the Grassmann expansion  of the wave function 
of the Universe, 
in the approach of Witten's supersymmetric quantum mechanics for all Bianchi
Class A cosmological models. In the Grassmannian calculus, one 
introduces fermionic variable without a direct physical meaning. Physical 
quantities of interest, such as $\rm \Psi^* \Psi$, require 
integration over the Grassmannian variables, obtaining in this way 
information about their influence on the unnormalized probability 
distributions. 

We begin with metrics of the form
\begin{equation}
\rm ds^2 = (N^2-N^jN_j)dt^2 -2N_i \, dt \, d\omega^i -e^{2\alpha(t)}
\, e^{2\beta(t)}\,_{ij} \omega^i \, \omega^j,
\end{equation}
where $\rm \alpha(t)$ is a scalar and $\rm \beta_{ij}$ is a $3\times 3$ 
matrix, and $\rm N(t)$ and $\rm N_i(t)$ are the lapse and shift functions, 
respectively. The one-forms $\rm \omega^i$ have specific
expression for any  particular Bianchi Universe \cite{RS}.

The Arnowitt-Deser-Misner (ADM) action has the form
\begin{equation}
\rm I= \int (P_x \, dx + P_y \, dy + P_\alpha \, d\alpha - N {\cal H}) dt ,
\label{action}
\end{equation}
where
\begin{equation}
{\cal H} =\rm  e^{-3\alpha}\left (-P_\alpha^2 + P_x^2 + P_y^2 + 
e^{4\alpha}\, V(x,y) \right) .
\label{hamiltonian}
\end{equation}
The potential term $\rm e^{4\alpha}\, V(x,y)= U(\alpha,x,y)$ is specific of 
the cosmological model under consideration, being related to  the 
function $\Phi(q^\mu)$ as follows \cite{OS}
($q^\mu=(\alpha,x,y)$).
\begin{equation}
\rm U(q^\mu)  = \eta^{\mu \nu} \frac{\partial \Phi}{\partial q^\mu} \,
\frac{\partial \Phi}{\partial q^\nu}  .
\label{HJ}
\end{equation}
This turns into the Einstein-Hamilton-Jacobi equation when we replace the 
momentum
$\rm P_\mu \rightarrow {\partial \Phi}/{\partial q^\mu}$ in 
(\ref{hamiltonian}), $\rm \eta^{\mu\nu}= diag(-1,1,1)$.

In this work we drop the factor $\rm e^{-3\alpha}$. This corresponds to 
fixing the factor ordering before factorizing the
WDW equation. The latter is got by replacing 
$\rm P_\mu \rightarrow -i\partial_{q^\mu}$ in (\ref{hamiltonian}).

The work is organized as follows. In Sec. II we give the factorization
of the WDW equation in terms of the supercharges operators and obtain
the so-called master equations for an auxiliar function 
$\rm f_\pm$ in general way for all Bianchi Class A cosmological models. 
In Sec. III we solve this master equation and present all bosonic components
of the wave function for the same models. In Sec. IV, we write a general
form for an unnormalized probability density, by imposing some soft of
boundary conditions on the bosonic components. Sec. V is devoted to 
conclusions

\section{Factorization of the WDW equation}

The WDW equation can be obtained as the bosonic sector of a 
super-Hamiltonian in superspace $\rm H_{ss}$  by employing  supercharge
operators. In the 3D case, the supercharges read
\begin{equation}
\rm Q = \psi^\mu \left[-P_\mu + i\frac{\partial \Phi}{\partial q^\mu} \right],
\label{supercharge1}
\end{equation}
and
\begin{equation}
\rm \bar Q = \bar \psi^\nu \left[-P_\nu -i 
\frac{\partial \Phi}{\partial q^\nu}  \right],
\label{supercharge2}
\end{equation}
with the following algebra for the variables $\psi^\mu$ and $\bar \psi^\nu$,
\begin{equation}
\left\{ \psi^\mu ,\bar \psi^\nu \right \} = \eta^{\mu\nu}, \qquad 
\left\{ \psi^\mu, \psi^\nu \right \} = 0, \qquad 
\left\{ \bar \psi^\mu,\bar \psi^\nu \right \} =0.
\end{equation}
 Using the representations $\psi^\mu=\eta^{\mu\nu} 
\partial/\partial \theta^\nu$ and $\bar \psi^\nu=\theta^\nu$, one finds the
superspace Hamiltonian in the form
\begin{equation}
\rm H_{ss}=\frac{1}{2} \left \{Q, \bar Q \right \} = {\cal H}_0 + 
\frac{\hbar}{2} \frac{\partial^2 \Phi}{\partial q^\mu \partial q^\nu}
\left[\bar \psi^\mu, \psi^\nu \right] ,
\label{superhamiltonian}
\end{equation}
where ${\cal H}_0=\square + U(q^\mu)$ is the standard WDW equation, 
$\square$ is the 3D d'Alambertian in the $\rm q^\mu$ coordinates with 
signature (- + +), $\left\{ \, , \, \right \}$ means the anti-commutator, 
and $\left[ \, , \, \right ]$, the commutator. Notice that the 
quantized super-Hamiltonian differs from the classical one ${\cal H}_0$ by 
a spin term, which vanishes in the classical limit.

The supercharges $\rm Q, \bar Q$ and the super-Hmiltonian satisfy the
following algebra
\begin{equation}
\frac{1}{2}\left\{Q,\bar Q\right\} = H_{ss}, \qquad 
\left\{H_{ss},Q \right\}=0, \qquad \left\{H_{ss}, \bar Q \right\}=0.
\end{equation}

Physical states are selected by the constraints

\begin{equation}
 Q \, \Psi =0,\qquad {\bar Q}\, \Psi = 0 , 
\label{physical}
\end{equation}
where the following 3D Grassmann representation of the $\Psi$ is used 
\begin{equation}
\rm \Psi= {\cal A}_+ + {\cal B}_\nu \theta^\nu 
+ \frac{1}{2}\epsilon_{\mu\nu\lambda} \, {\cal C}^\lambda \, \theta^\mu \,
\theta^\nu + {\cal A}_- \, \theta^0 \, \theta^1 \, \theta^2, 
\label{wavefunction}
\end{equation}
 $\mu, \nu, \lambda$ running over $0,1,2.$

The following ansatz  
\begin{equation}
{\cal B}_\mu ={\rm  \frac{\partial f_+}{\partial q^\mu}\, e^{-\Phi}} 
\end{equation}
in (\ref{physical}) and (\ref{wavefunction}), leads to the master equation
for the auxiliary function $\rm f_+$
\begin{equation}
\rm \square f_+ - 2 \eta^{\mu \nu} \frac{\partial \Phi}{\partial q^\mu} 
\frac{\partial f_+}{\partial q^\nu}=0.
\label{master1}
\end{equation}
In addition, it is possible to show that 
$\frac{1}{2}\epsilon_{\mu\nu\lambda} \, 
{\cal C}^\lambda \, \theta^\alpha \, \theta^\mu \, \theta^\nu={\cal C}^\alpha
\theta^0 \theta^1 \theta^2$
and employing the ansatz 
\begin{equation}
{\cal C}^\alpha = {\rm \eta^{\alpha \mu}\frac{\partial f_-}{\partial q^\mu}\, 
e^{\Phi}}
\end{equation}
we obtain the second master equation of the form
\begin{equation}
\rm \square f_- + 2 \eta^{\mu \nu} \frac{\partial \Phi}{\partial q^\mu} 
\frac{\partial f_-}{\partial q^\nu}=0.
\label{master2}
\end{equation}
Thus, (\ref{master1}) and (\ref{master2}) can be rewritten  as
\begin{equation}
\rm \square f_\pm \mp 2 \eta^{\mu \nu} \frac{\partial \Phi}{\partial q^\mu} 
\frac{\partial f_\pm}{\partial q^\nu}=0.
\label{master3}
\end{equation}

The solutions for the other functions are
\begin{equation}
{\cal A}_\pm =q_\pm \, e^{\mp \Phi}
\end{equation}
where $\rm q_\pm$ are integration constants.

In the following, we give the solution of (\ref{master3}), using as toy 
models the Bianchi Class A cosmological models \cite{OS}


\section{Quantum solutions for Bianchi Class A models}

To solve  (\ref{master3}) it is necessary to know the potentials 
$\rm U(q^\mu)$ (or $\Phi(q^\mu)$) for the Bianchi Class A cosmological 
models. These are given in  table I \cite{OS}

\begin{center}
\begin{tabular}{|l|l|l|} \hline
Bianchi type  &\hglue 3.5cm Potential U &\hglue 1.7cm $\Phi$ \\  \hline
I  &  0  & 0   \\  \hline 
II & $\rm \frac{1}{3}\, e^{4\beta_1}$ & $\rm \pm \frac{1}{6}\, e^{2\beta_1}$
  \\ \hline
$\rm VI_{h=-1}$  & $\frac{4}{3}\, e^{2(\beta_1+\beta_2)}$ &  
$\rm \pm \frac{1}{6}\, \left[2(\beta_1 - \beta_2)\right] e^{\beta_1+\beta_2}$
   \\ \hline
$\rm VII_{h=0}$  &$ \frac{1}{3}\,\left[e^{4\beta_1} + e^{4\beta_2}-
2 e^{2(\beta_1+\beta_2)}\right]$ &  
$\rm \pm \frac{1}{6}\, \left[e^{2\beta_1}+e^{2\beta_2}   \right]$
   \\  \hline
$\rm VIII$  & $\frac{1}{3}\,\left[e^{4\beta_1} + e^{4\beta_2}+e^{4\beta_3} 
-2 e^{2(\beta_1+\beta_2)}+2 e^{2(\beta_1+\beta_3)}+2 e^{2(\beta_2+\beta_3)}
\right]$ 
& $\rm \pm \frac{1}{6}\, \left[e^{2\beta_1}+e^{2\beta_2} -e^{2\beta_3} \right]$
  \\ \hline 
$\rm IX$  & $\frac{1}{3}\,\left[e^{4\beta_1} + e^{4\beta_2}+e^{4\beta_3} 
+2 e^{2(\beta_1+\beta_2)}+2 e^{2(\beta_1+\beta_3)}+2 e^{2(\beta_2+\beta_3)}
\right]$ 
& $\rm \pm \frac{1}{6}\, \left[e^{2\beta_1}+e^{2\beta_2} +e^{2\beta_3} 
\right]$
  \\ \hline 
\end{tabular} 
{\small Table I\\
Potential U and superpotential $\rm \Phi$ for all Bianchi Class A 
models, where $\rm \beta_1=\alpha + x + \sqrt{3} \,  y, \, 
\beta_2=\alpha + x - \sqrt{3}\,  y$ and $\rm \beta_3=\alpha -2 x $. }   
\end{center}

Once $\rm f_\pm$ are obtained, all the bosonic components that appear
in the Grassmann expansion of the wave function (\ref{wavefunction})
can be determined as follows. The ansatz
\begin{equation}
\rm f_\pm(q^\mu) = W_\pm(q^\mu) \, e^{\pm \Phi(q^\mu)}.
\label{ansatz1}
\end{equation}
turns (\ref{master3}) into
\begin{equation}
\rm \square W_\pm = \left[(\nabla \Phi)^2 \mp \square \Phi  \right ]\, 
W_\pm \, ,
\end{equation}
whose solutions are
\begin{equation}
\rm W_\pm(q^\mu) = \beta_\pm e^{m_\mu t^\mu}\, e^{\mp \Phi},
\label{solution}
\end{equation}
where $\rm \beta_\pm$ are  integration constants, $\rm m_\mu=(m_1,m_2,m_3)$ 
is a vector of null measure (i.e. $\rm -m_1^2+m_2^2+m_3^2=0$) and 
$\rm t^\mu=(-\alpha,x,y)$.

Thus, the solutions for  $\rm f_\pm$ are given by
\begin{equation}
\rm f_\pm = \beta_\pm e^{m_\mu t^\mu}.
\end{equation}
In the Table II, we write the corresponding $\rm m_\mu t^\mu$ terms for 
all Bianchi Class A models.

  \begin{center}
 \begin{tabular}{|l|c|c|} \hline
Bianchi type  & $\rm m_\mu t^\mu=-m_{1_\pm} \, \alpha+m_{2_\pm}\, x + 
m_{3_\pm}\,y$ & $\rm m_\mu $        \\  \hline
I  & $\rm -m_{1_\pm}\, \alpha+m_{2_\pm}\, x + m_{3_\pm}\, y$&$ 
(m_{1_\pm},m_{2_\pm},m_{3_\pm})$   \\  \hline 
II & $\rm b_\pm (2\alpha -x + \sqrt{3}\,  y)$&
$\rm b_\pm(-2,-1,\sqrt {3})$\\ \hline
$\rm VI_{h=-1}$  & $\rm  b_\pm \, (\alpha + x)$& 
$\rm b_\pm (-1,1,0)$           \\ \hline
$\rm VII_{h=0}$  &$ \rm  b_\pm \, (\alpha + x)  $ &
$\rm b_\pm (-1,1,0)$  \\  \hline
$\rm VIII$  & 0 & $(0,0,0)$  \\ \hline 
$\rm IX$  &   0 & $(0,0,0)$  \\ \hline 
\end{tabular}   
     
{\small Table II\\
$\rm m_\mu t^\mu$ and $\rm m_\mu$ terms for all Bianchi Class 
A models.}
  \end{center}

Moreover, the solutions for all bosonic components of the wave  function 
are

\begin{eqnarray}
{\cal A}_\pm &=& {\rm q_\pm \, e^{\mp \Phi}} \, \nonumber\\
{\cal B}_0&=& {\rm -m_{1_+} \, f_+ e^{-\Phi}}, 
\qquad {\cal C}^0={\rm m_{1_-} \, f_- \, e^{\Phi} },\label{bosonic} \\
{\cal B}_1&=& {\rm m_{2_+}\, f_+ \, e^{-\Phi} }, \qquad 
{\cal C}^1={\rm m_{2_-}\, f_- \, e^{\Phi} } , \nonumber\\
{\cal B}_2&=& {\rm m_{3_+}\, f_+ \, e^{-\Phi} } , \qquad 
{\cal C}^2={\rm m_{3_-}\, f_- \, e^{\Phi} }, \nonumber
\end{eqnarray}
where $\rm q_\pm  \mbox{and} b_\pm$ are arbitrary constants and 
$\rm m_{i_\pm}$ are constants given in  table II.

\section{The unnormalized probability density $|\Psi|^2$}

In this section we follow ref. \cite{OSB}  to get $|\Psi|^2$
for the wave function given by (\ref{wavefunction}). Integrating over the 
Grassmann variables $\theta^\mu$ \cite{Faddeev}, we are led to the 
following  result 
\begin{equation}
\rm |\Psi|^2={\cal A}_+^* \, {\cal A}_+ + {\cal B}_0^* \, {\cal B}_0
+{\cal B}_1^* \, {\cal B}_1 + {\cal B}_2^* \, {\cal B}_2
 + {\cal C}^{0*} \, {\cal C}^0 + {\cal C}^{1*} \, {\cal C}^1
+ {\cal C}^{2*} \, {\cal C}^2 + {\cal A}_-^* \, {\cal A}_- \, .
\label{density}
\end{equation}
Assuming that the constants  appearing in (\ref{density}) are real,
we obtain
\begin{equation}
\rm |\Psi|^2= \left[ q_+^2 +  (m_{1_+}^2 +m_{2_+}^2+m_{3_+}^2)\, f_+^2 
\right]\,  e^{-2\Phi} +
\left[ q_-^2 + (m_{1_-}^2 +m_{2_-}^2+m_{3_-}^2)\, f_-^2  \right]\, 
e^{2\Phi} .
\label{density1}
\end{equation}

 For example, for the Bianchi type II models, we obtain from 
(\ref{density1})
\begin{equation}
\rm |\Psi_{II}|^2= \left( q_+^2 +  8b_+^2\, f_+^2 \right)\,  e^{-2\Phi} +
\left( q_-^2 + 8 b_-^2\, f_-^2  \right)\, e^{2\Phi},
\label{density2}
\end{equation}
which is Eq. (4.2) in \cite{OSB}.

Now, by imposing the  some sort of boundary conditions, only one kind of 
terms will remain in the wave function \cite{HP}. There are many 
possibilities, one is to demand that $\rm \Psi$ does not diverge for 
$\rm |x|$, $\rm |y| \rightarrow \infty$ at fixed $\alpha$.  In such a case 
only the terms $\rm e^{-\Phi}$ will remain in (\ref{density1}). 
 
Figures 1 and 2, display the behavior of the  probabilitity density
$\rm |\Psi|^2$ for the Bianchi type VI$_{h=-1}$ and VII$_{h=0}$ when
the parameter $\rm b_+=-1$, respectively. The $|\Psi|^2$ for the same
models when $\rm b_+=+1$ are plotted in Figures 3 and 4. 
The $|\Psi|^2$ behavior for the Bianchi type II and IX, has been given 
in \cite{OSB} and \cite{Graham1}, respectively.

\section{Conclusion}
Our main goal in this work, was to obtain a general expression for all bosonic
functions that appear in the expansion  of the wave function of the Universe 
using the approach of Witten's supersymmetric quantum mechanics. This has 
been done for all Bianchi Class A cosmological models. In addition, we found
the general form for the unnormalized probabilitity density, 
(\ref{density1}), for any Bianchi Class A cosmological model. When we apply 
boundary conditions on the wave function, the unnormalized probability 
density is reduced to
\begin{equation}
\rm |\Psi|^2= \left[ q_+^2 +  (m_{1_+}^2 +m_{2_+}^2+m_{3_+}^2)\, f_+^2 
\right]\,  e^{-2\Phi} .
\label{density3}
\end{equation}

At this step, the main contribution to $|\Psi|^2$ will depend on the
values of the parameter $\rm b_+$ that occurs in the function $\rm f_+$ and
on the range used  for the coordinates $\alpha$ and $\rm (x,y)$.

\section{\bf Acknowledgments}
We would like to thank H. Rosu for valuable hints.
This research was supported by  CONACyT. ERM  was supported by a CONACYT 
student fellowship.

\newpage

\centerline{
\epsfxsize=260pt
\epsfbox{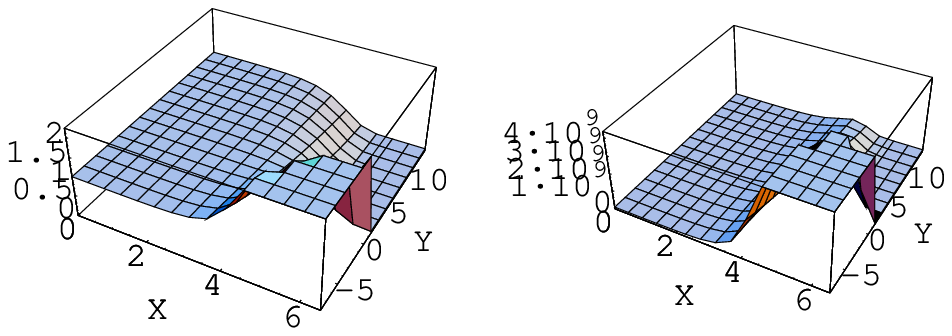}}
\vskip 4ex
\begin{center}
{\small {Fig. 1}\\ 
At the left we show $\rm |\Psi_{VI}|^2= q^2_+ \, e^{-2\Phi}$, for 
the first pure bosonic term, when $\rm q_+=1, \alpha=-6$. At the right,
 the main contribution to the unnormalized probability density
due to the  ${\cal B}_\mu$ terms, when $\beta_+=1, \, b_+=-1, \, \alpha=-6$.
 }
\end{center}

\centerline{
\epsfxsize=260pt
\epsfbox{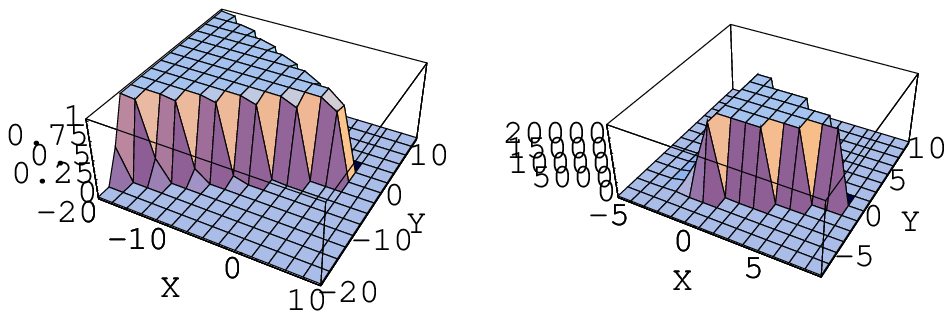}}
\vskip 4ex
\begin{center}
{\small {Fig. 2}\\
At the left we show $\rm |\Psi_{VII}|^2= q^2_+ \, e^{-2\Phi}$, for 
the first pure bosonic term when $\rm q_+=1, \alpha=-6$. At the right, the 
main contribution to the unnormalized probability density due to the
${\cal B}_\mu$ terms, when $\beta_+=1,\,b_+=-1, \, \alpha=-6$.
 }
\end{center}
\newpage
\centerline{
\epsfxsize=260pt
\epsfbox{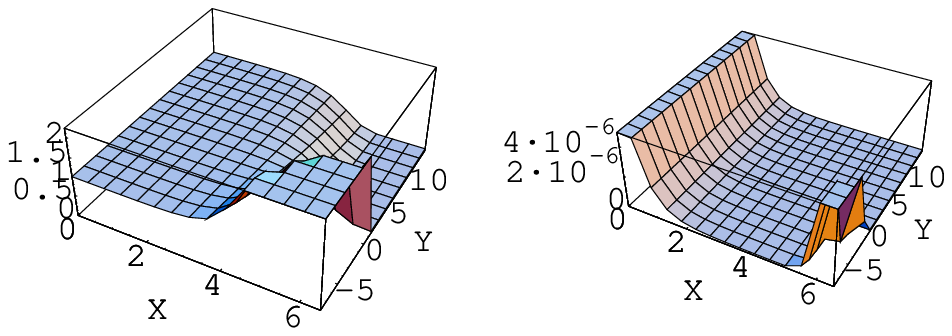}}
\vskip 4ex
\begin{center}
{\small {Fig. 3}\\ 
At the left we show $\rm |\Psi_{VI}|^2= q^2_+ \, e^{-2\Phi}$, the main 
contribution to the unnormalized probability density due to the first 
pure bosonic term, when $\rm q_+=1, \alpha=-6$. At the right,  
the corresponding contribution of the 
${\cal B}_\mu$ terms, for $\beta_+=1,\,b_+=1,\, \alpha=-6$.
 }
\end{center}

\centerline{
\epsfxsize=260pt
\epsfbox{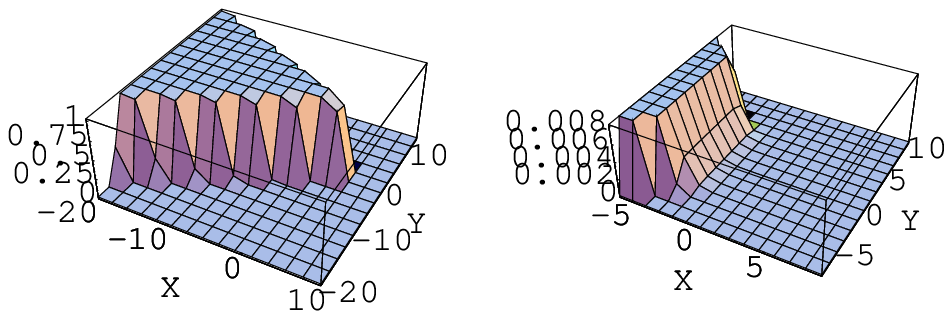}}
\vskip 4ex
\begin{center}
{\small {Fig. 4}\\
At the left we show $\rm |\Psi_{VII}|^2= q^2_+ \, e^{-2\Phi}$, the main 
contribution to the unnormalized probability density due to the first 
pure bosonic term, when $\rm q_+=1, \alpha=-6$. At the right,  
the corresponding contribution of the 
${\cal B}_\mu$ terms, for $\beta_+=1,\,b_+=1,\, \alpha=-6$.
 }
\end{center}

\end{document}